\documentclass[aps,prx,twocolumn,showpacs,psfig,superscriptaddress,longbibliography]{revtex4-1}
\usepackage{amssymb}
\usepackage{algpseudocode}
\usepackage{algorithm}
\usepackage{amsmath}
\usepackage{braket}
\usepackage{booktabs}
\usepackage{chngcntr}
\usepackage{color}
\usepackage{comment}
\usepackage{dsfont}
\usepackage{etoolbox}
\usepackage{epstopdf}
\usepackage[T1]{fontenc}
\usepackage{float}
\usepackage{graphicx}
\usepackage[colorlinks=true,linkcolor=blue,citecolor=blue,urlcolor=blue]{hyperref}
\usepackage[latin9]{inputenc}
\usepackage{indentfirst}
\usepackage{latexsym,bm,euscript}
\usepackage{mathrsfs}
\usepackage{multirow}
\usepackage{natbib}
\usepackage{soul}
\usepackage{subfigure}
\usepackage{times}
\usepackage{titlesec}
\usepackage{tikz}
\usepackage{txfonts}
\usepackage{xspace}
\usepackage{xfrac}

\def\lsb#1{\left(#1\right)}              
\def\lmb#1{\left[#1\right]}              
\def\llb#1{\left\{#1\right\}}            

\def\beqn{\begin{eqnarray}}
\def\eeqn{\end{eqnarray}}
\def\beq{\begin{equation}}
\def\eeq{\end{equation}}
\def\pd#1#2{\frac{\partial{#1}}{\partial{#2}}}
\def\squad{\hskip0.5em\relax}

\newcommand{\Sec}[1]{Sec.~\ref{#1}}

\newcommand{\Eq}[1]{Eq.~\eqref{#1}}

\newcommand{\Fig}[1]{Fig.~\ref{#1}}

\def\dTRG{$\partial$TRG\xspace}

\graphicspath{{./Fig/}}

\begin{document}
\title{Automatic Differentiation for Second Renormalization of Tensor Networks}

\author{Bin-Bin Chen}
\affiliation{School of Physics, Key Laboratory of Micro-Nano Measurement-Manipulation and Physics (Ministry of Education), 
Beihang University, Beijing 100191, China}
\affiliation{Physics Department, Arnold Sommerfeld Center for Theoretical Physics,
and Center for NanoScience, Ludwig-Maximilians-Universit\"at,
Theresienstrasse 37, 80333 Munich, Germany}

\author{Yuan Gao}
\affiliation{School of Physics, Key Laboratory of Micro-Nano Measurement-Manipulation and Physics (Ministry of Education), 
Beihang University, Beijing 100191, China}

\author{Yi-Bin Guo}
\affiliation{Institute of Physics, Chinese Academy of Sciences, P.O. Box 603, Beijing 100190, China}
\affiliation{School of Physics, University of Chinese Academy of Sciences, Beijing 100049, China}

\author{Yuzhi Liu}
\affiliation{Department of Physics, Indiana University, Bloomington, Indiana 47405, USA}

\author{Hui-Hai Zhao}
\affiliation{Alibaba Quantum Laboratory, Alibaba Group, Beijing, China}

\author{Hai-Jun Liao}
\affiliation{Institute of Physics, Chinese Academy of Sciences, P.O. Box 603, Beijing 100190, China}
\affiliation{Songshan Lake Materials Laboratory, Dongguan, Guangdong 523808, China}

\author{Lei Wang}
\affiliation{Institute of Physics, Chinese Academy of Sciences, P.O. Box 603, Beijing 100190, China}
\affiliation{Songshan Lake Materials Laboratory, Dongguan, Guangdong 523808, China}

\author{Tao Xiang}
\affiliation{Institute of Physics, Chinese Academy of Sciences, P.O. Box 603, Beijing 100190, China}
\affiliation{School of Physics, University of Chinese Academy of Sciences, Beijing 100049, China}
\affiliation{Kavli Institute for Theoretical Sciences, University of Chinese Academy of Sciences, Beijing 100190, China}

\author{Wei Li}
\email{w.li@buaa.edu.cn}
\affiliation{School of Physics, Key Laboratory of Micro-Nano Measurement-Manipulation and Physics (Ministry of Education), 
Beihang University, Beijing 100191, China}

\author{Z. Y. Xie}
\email{qingtaoxie@ruc.edu.cn}
\affiliation{Department of Physics, Renmin University of China, Beijing 100872, China}

\begin{abstract} 
Tensor renormalization group (TRG) constitutes an important methodology for accurate simulations of strongly correlated lattice models. Facilitated by the automatic differentiation technique widely used in deep learning, we propose a uniform framework of differentiable TRG (\dTRG) that can be applied to improve various TRG methods, in an automatic fashion. \dTRG systematically extends the essential concept of second renormalization [PRL 103, 160601 (2009)] where the tensor environment is computed recursively in the backward iteration. Given the forward TRG process, \dTRG automatically finds the gradient of local tensors through backpropagation, with which one can deeply ``train'' the tensor networks. We benchmark \dTRG in solving the square-lattice Ising model, and demonstrate its power by simulating one- and two-dimensional quantum systems at finite temperature. The global optimization as well as GPU acceleration renders \dTRG a highly efficient and accurate manybody computation approach.
\end{abstract}

\date{\today}
\maketitle

\textit{Introduction.---}  
In the investigation of strongly correlated quantum states and materials, 
tensor renormalization group (TRG) constitutes a thriving field that is
playing an increasingly important role recently.  In the diverse family of TRG approaches, 
there include the coarse-graining TRG \cite{Levin.m+:2007:TRG,TEFR-PRB2009},  
higher-order TRG (HOTRG) \cite{Xie2012}, and tensor network renormalization \cite{TNR-PRL2015,TNR-MERA,LoopTRG-PRL2017,PosTRG-PRL2017}. 
They have been put forward to evaluate classical statistical systems as well as expectation values out of two-dimensional (2D) tensor network states \cite{Jiang2008}. 
There are also TRG methods developed to simulate $d$-dimensional quantum lattice models at finite temperature \cite{Li.w+:2011:LTRG,Czarnik.p+:2012:PEPS,Dong.y+:2017:BiLTRG,Kshetrimayum2019,Czarnik.p+:2015:PEPS,Luca2018,Czarnik.P2019b,Chen2018,Li2019}, whose Euclidean path integral constitutes a ($d+1$)-dimensional worldsheet. 

\begin{figure}[!tbp]
\includegraphics[angle=0,width=1\linewidth]{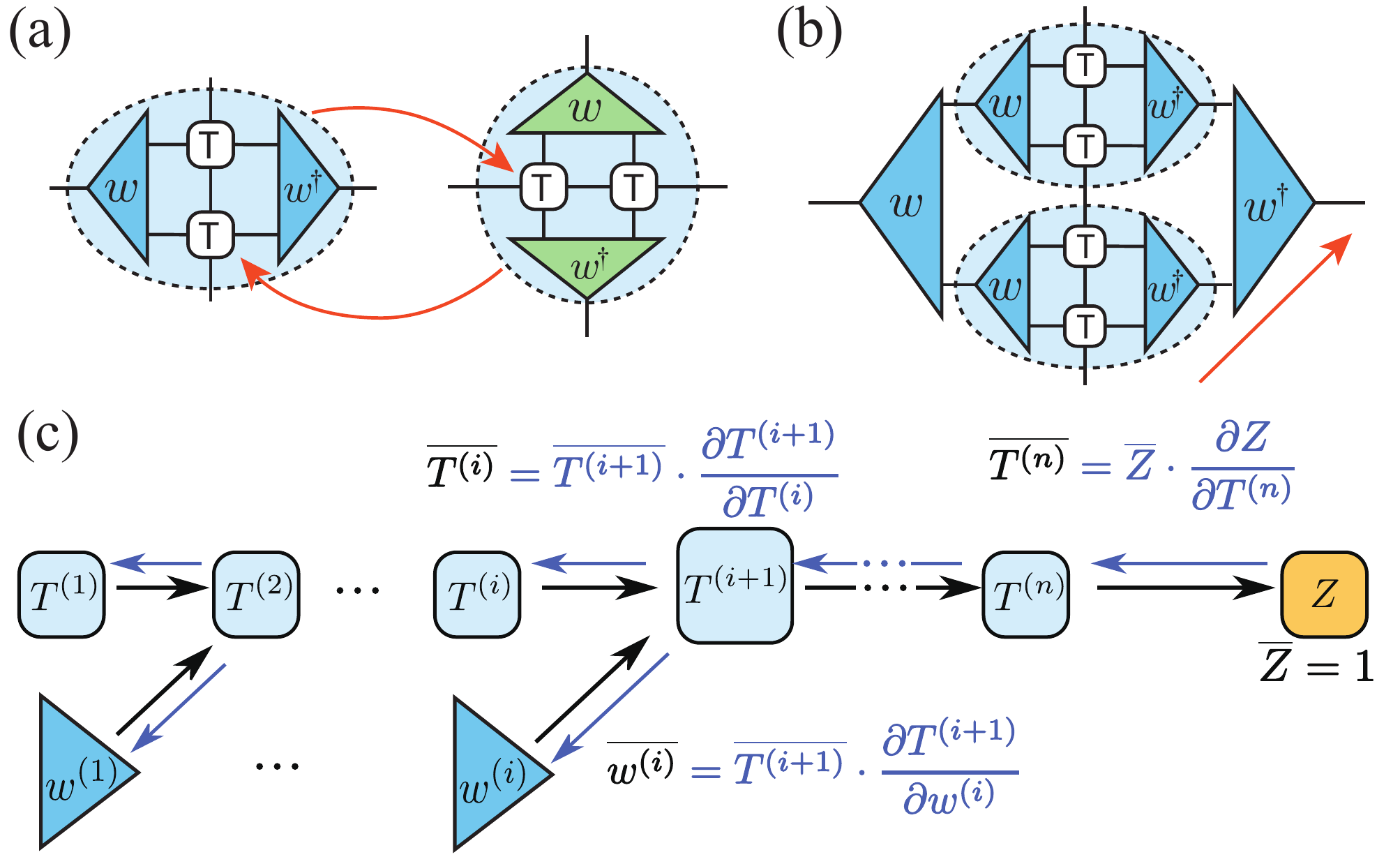}
\caption{(Color online) (a) shows a TRG step where the scale transformations $w$ are introduced along both directions, while only vertical renormalization is involved in (b), which eventually compresses the tensor network into a 1D structure that can then be contracted exactly. 
(c) plots the computational graph of the forward TRG process as well as the backpropagation.}
\label{Fig:dTRG}
\end{figure}

In the course of TRG process, environment of local tensors should 
be taken into account for conducting a precise truncation,
through, e.g., isometric renormalization transformations 
in the tensor bases. This can be traced back to 
the renowned density matrix renormalization group \cite{White1992}, where the effects of environment 
are reflected in the reduced density matrix of ``system" subblock. 
For generic tensor networks, second renormalization group (SRG) 
has been proposed to improve the process of tensor renormalization
\cite{Xie2009,SRG-PRB2010,Xie2012,HOSRG-PRB2016}. 
In SRG, the environment of local tensors is computed
recursively, between different scales of a hierarchical network, 
with which a global optimization is feasible. 

Recently, profound interplay between deep learning and tensor network algorithms has raised great interest \cite{Carleo2017, Carrasquilla2017,Nieuwenburg2017,Stoudenmire2016,Foreman2018,Han2018,Guo2018,SHLi2018,Koch-Janusz2018}. Among others, 
the differentiable programming is of particular interest for tensor networks. 
Given the computational graph generated in forward process, 
the gradient of corresponding variables can be calculated through the 
chain rule of derivatives in the backpropagation, with which the neural 
network can be deeply trained. The automatic calculation of gradients can 
be obtained within machine precision, and with same computational 
complexity of forward process. Very recently, this idea of differentiable 
programming that exploits a gradient based optimization, has been introduced to optimize tensor networks \cite{Liao2019}. 

In this work, we regard the renormalization transformation as input parameters of the TRG program, and point out that the SRG backward iteration bears a correspondence with the backpropagation algorithm in differentiable programming. Inspired by this substantial connection, we turn the idea of SRG into a generalized versatile framework, i.e., differentiable TRG (\dTRG). 

In \dTRG, the forward TRG process is made fully differentiable, and the renormalization transformations are optimized globally and automatically through the backpropagation. We apply \dTRG to simulate thermal equilibrium states at finite temperature, and achieve significantly improved accuracy over previous methods \cite{Li.w+:2011:LTRG,Chen2018}. The efficiency is demonstrated by implementing \dTRG with PyTorch \cite{Pytorch,pytorch}, which facilitates the GPU computing and shows a high performance of about 40 times acceleration over a single CPU core.

\textit{Correspondence between SRG and backpropagation.}--- Backpropagation is a widely used method for training deep neural networks \cite{Hinton-Nature1986,Parker-LLR1985,LeCun-IEEE1989,LeCun-Nature2015}, where the gradients of parameters can be computed through a reverse-mode automatic differentiation \cite{SM}. 
On the other hand, SRG plays a very similar role in tensor-network algorithms as the backpropagation. To be specific, as shown in Figs.~\ref{Fig:dTRG}(a,c), a hierarchical tensor network can be constructed by piling up a series of isometric RG transformations $\llb{w^{(i)}}$, with $i = 1, 2, ..., n $ for each layer. In a well-designed TRG program, the input tensors $\llb{w^{(i)}}$ are successively applied to the tensors $T$ and the output can be a tensor trace in general, e.g., partition function of a statistical system.

SRG takes the job to further optimize the renormalization
transformations $\llb{w^{(i)}}$ in the backward iteration, 
by making use of  the environment. 
An adjoint tensor of $w^{(i)}$ at scale $i$ is defined as the gradient $
\overline{w^{(i)}}$ ($\equiv \frac{\partial Z}{\partial w^{(i)}}$), which can be related to the environment through
\beq
E_{w}^{(i)} = \frac{1}{N_i} \overline{w^{(i)}},  \label{EGrad}
\eeq
where $N_i$ denotes the times $w^{(i)}$ appearing in the network, and 
$Z$ the tensor trace to be maximized. It directly follows from 
Eq.~(\ref{EGrad}) that the recursive relations used in the backward 
iteration to determine $\llb{E^{(i)}}$ in SRG, can be recasted into the 
derivative chain rule form, as depicted in Fig.~\ref{Fig:dTRG}(c). Remind 
that the multiplications there with Jacobians ${\partial T^{(i+1)}}/{\partial 
T^{(i)}}$ and ${\partial T^{(i+1)}}/{\partial w^{(i)}}$, etc, are conducted 
implicitly. They constitute sequences of tensor contractions exactly 
equivalent to the recursive tensor contractions in SRG \cite{SM}. 

\begin{figure}[!b]
\includegraphics[angle=0,width=1\linewidth]{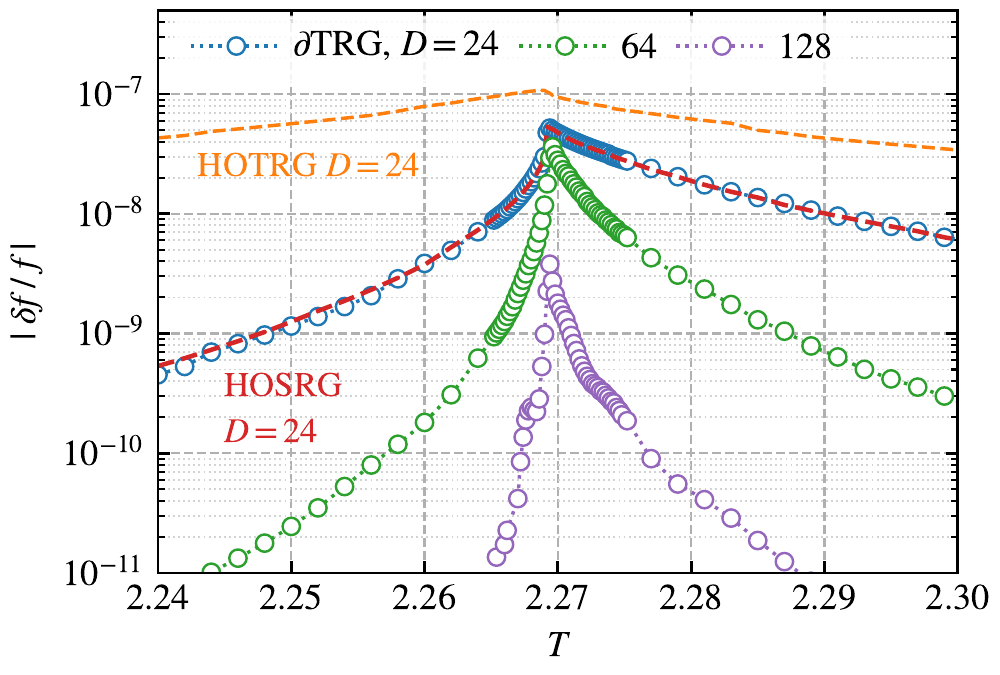}
\caption{(Color online) The relative errors of free energy $|\delta f/f|$ in the vicinity of critical temperature $T_c$, obtained from HOTRG, HOSRG, and \dTRG calculations. We perform $n_i=5$--10 iterations within each MERA update, and the total number of sweeps $n_s$ ranges from a few iterations to a few hundreds, depending on the specific temperature and scheme, until the relative errors reach a convergence criterion of $\epsilon=10^{-4}$. The $D=24$ \dTRG calculations follows the scheme in \Fig{Fig:dTRG}(a), while $D=64$ and 128 cases the scheme in \Fig{Fig:dTRG}(b).} 
\label{Fig:cdErr}
\end{figure}

\textit{Differentiable tensor renormalization group.}--- Being aware of the intimate relation between the backpropagation and SRG, we now extend the latter to a more general and flexible framework, \dTRG, with the help of well-developed automatic differentiation packages \cite{Wright-NO2006}, e.g., autograd \cite{autograd} and PyTorch \cite{Pytorch,pytorch}. With these facilities, \dTRG can record all operations performed on the input variables (tensors), and compute the derivatives [e.g.~\Eq{EGrad}] automatically in the backward iterations, with which the parameters $\llb{w^{(i)}}$ can be optimized. Not limited within the original proposals \cite{Xie2009,Xie2012}, the idea of SRG can be applied to various TRG schemes through the framework of \textit{differentiable programming}. Below we consider two different \dTRG schemes following the HOTRG \cite{Xie2012} and exponential TRG (XTRG) \cite{Chen2018}, as shown in  Figs.~\ref{Fig:dTRG}(a) and (b), respectively.

Once the environment $E_w^{(i)}$ is obtained, one can optimize $w^{(i)}$ with resorting to, e.g., 
standard quasi-Newton optimization method \cite{SM}, or quasi-optimal schemes through  
tensor decompositions of $E_w^{(i)}$ \cite{Xie2009, Xie2012}. 
Remind that a tensor decomposition scheme that keeps $w^{(i)}$ isometric 
has been developed in the context of multi-scale entanglement renormalization ansatz (MERA) 
algorithm \cite{Vidal2007,Evenbly2009}, which are mainly adopted in the simulations below. 
MERA update involves a singular value decomposition (SVD) $E_w = U S V^\dagger$ and 
a replacement $w=U V^{\dagger}$, which maximizes the cost function 
$Z =$ Tr $(E_w \cdot w)$, with $O(D^4)$ time complexity. 
Here $D$ is the geometric bond dimension of a tensor.

Due to the intrinsic nonlinearility (in $w$) in the optimization problem, 
$n_i$ inner iterations are introduced in a single step of MERA update ($n_i=5$--$10$ in practice). 
Moreover, thanks to the convenient access to $E_w^{(i)}$, 
in \dTRG we can deeply optimize the tensor network via sweep optimizations. 
In practice, we scan from inner to outer layers $n_s$ times until the results converge, 
thus assuring a highly accurate global update of $\llb{w^{(i)}}$ tensors. 

\textit{\dTRG of 2D Ising model.---}  As a first demonstration, we apply \dTRG, 
with two specific implementations in Figs.~\ref{Fig:dTRG}(a,b), 
to solve the classical Ising model on the square lattice. Following the standard procedure, 
we can write down a square tensor-network representation consisted of rank-4 tensors $T$, 
whose TRG contraction results in the partition function $Z$ \cite{SM}.  

In \Fig{Fig:dTRG}(a), after $n$ steps of renormalizations, we obtain a single tensor representing the whole system of $2^{n} \times 2^{n}$ sites (in practice $n=25$ guarantees the thermodynamic limit), whose self-contraction leads to the partition function $Z$. On the other hand, after $n$ steps of renormalization, one arrives at an effective 1D system, whose complete contraction also leads to an accurate measure of the partition function. 

In \Fig{Fig:cdErr}, we show the accuracies of \dTRG implementations, together with the HOTRG and HOSRG data for comparisons. Owing to the sweep update, \dTRG leads to errors clearly smaller than those of HOTRG, while achieving, as expected, the same accuracies as HOSRG \cite{footnote}. 

The two schemes of \dTRG in Figs.~\ref{Fig:dTRG}(a) and (b) have different computational costs. 
The latter is considerably less resource-demanding, i.e., $O(D^4)$ in computational time, while it is $O(D^7)$ in \Fig{Fig:dTRG}(a). 
The memory costs are also dramatically different, i.e., $O(D^5)$ for \Fig{Fig:dTRG}(a) and $O(D^3)$ for (b). 
Therefore, we can push the \dTRG simulations in \Fig{Fig:dTRG}(b)
with bond states up to $D=128$, reaching much higher precision as shown in \Fig{Fig:cdErr}.
From the comparison shown in \Fig{Fig:cdErr}, as well as other considerations, 
we chose the \dTRG scheme in \Fig{Fig:dTRG}(b) to simulate quantum models as presented below.

\begin{figure}[!tbp]
\includegraphics[angle=0,width=1\linewidth]{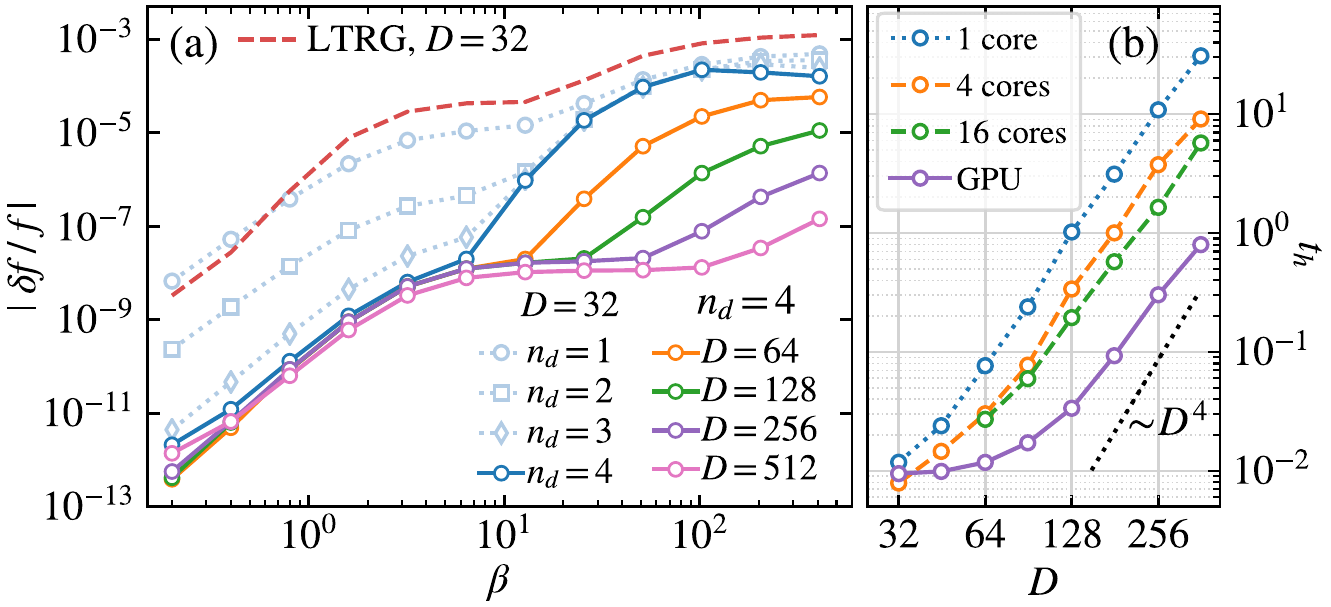}
\caption{(Color online) 
(a) Comparisons of relative errors of free energy between linearized TRG (LTRG) and $\partial$TRG. In the initial $\rho(\tau)$ of $\partial$TRG, $\tau\simeq 5\times10^{-5}$, which is also used as the Trotter slice in the LTRG calculations. The results are shown with optimization depth $n_d \leq 4$, $n_i=10$ in a single MERA update, and overall sweep iterations $n_{\rm{s}}=3$. 
(b) Comparisons of the computational walltime of $\partial$TRG on GPU and CPU with up to 16 cores, benchmarked on the infinite XY chain. The calculations are carried out on Nvidia Tesla V100 GPU and Intel Xeon 6230 CPU, with retained bond dimensions up to $D = 360$. The dashed line depicts the scaling $t_h \sim D^4$.
}
\label{Fig:InfXYC}
\end{figure}

\textit{Infinite quantum XY chain.---} 
Now we employ $\partial$TRG to simulate the 
exactly solvable quantum XY chain 
\begin{equation}
H_{XY} = \sum_{\langle i,j \rangle} S_i^x S_j^x + S_i^y S_j^y. 
\end{equation}
We start with preparing the density matrix $\rho(\tau)$ through (second order) Trotter-Suzuki decomposition \cite{SM}. 
A very small imaginary-time step $\tau$ is chosen to ensure that Trotter errors are negligible. 
Given the matrix product operator (MPO) representation of $\rho(\tau)$, 
we proceed to cool down the system exponentially fast, with the \dTRG algorithm shown in \Fig{Fig:dTRG}(b). 
The results are shown in \Fig{Fig:InfXYC}(a), where the relative error $|\delta f/f|$ curves rise up from very small values 
at high temperature and increase monotonically as $T$ decreases. 

In \Fig{Fig:InfXYC}(a), benchmarking with the analytical solution \cite{Tu-CFT2017, Chen-CFT2017, SM}, we compare the relative errors $|\delta f/f|$ between \dTRG and LTRG, 
where the latter follows a cooling procedure linear in $\beta (\equiv 1/T)$ \cite{Li.w+:2011:LTRG}. 
It is observed that $\partial$TRG with depth $n_d=1$ (i.e., optimizing exclusively the current layer in the course of cooling) 
already outperforms LTRG in both efficiency and accuracy. By sweeping into $n_d$ (up to 4) layers, the 
accuracy is found to improve continuously in the relatively high to intermediate temperature regime due to better optimization. 
At low temperature, on the other hand, the enhancement of accuracy is marginal due to the limited expressibility 
of the tensor network with a given bond dimension $D=32$. Therefore, we show also in \Fig{Fig:InfXYC}(a) 
the results of larger bond dimensions (up to 512), with a fixed depth $n_d=4$. There we observe that $|\delta f/f|$ 
decreases monotonically and attains very high accuracy, with relative error 
$\sim 10^{-7}$ at low temperature (down to $\beta \simeq 400$).

\textit{GPU acceleration.---} We implement $\partial$TRG with the PyTorch library, 
and take advantage of GPU computing to significantly accelerate the simulations. 
In \Fig{Fig:InfXYC}(b), we show the elapsed hours $t_h$ versus $D$ in the simulations 
of infinite XY chain on GPU and CPU, respectively. To quantify the speedup, $t_h$ is monitored at $\beta=12.8$, 
where the computation time falls well into a logarithmic scaling regime vs. $\beta$, i.e., $t_h \propto \ln{\beta}$ \cite{SM}.

From \Fig{Fig:InfXYC}(b), we observe approximately 40 times GPU acceleration (for $D=360$ calculations), as compared to single core CPU calculations, 
and over 7 times speedup to the 16-core parallel job. Moreover, in \Fig{Fig:InfXYC}(b), the $t_h$ curves show algebraic scaling vs. $D$, i.e., $t_h \sim D^\gamma$, 
for sufficiently large $D$ where $\gamma$ values are found slightly less than 4. These appealing benchmarks, together with previous tests in Ref.~\cite{Milsted2019}, 
suggest that GPU acceleration indeed constitutes a very promising technique to be fully explored in quantum manybody computations, 
particularly in tensor network simulations.

\begin{figure}[!tbp]
\includegraphics[angle=0,width=1\linewidth]{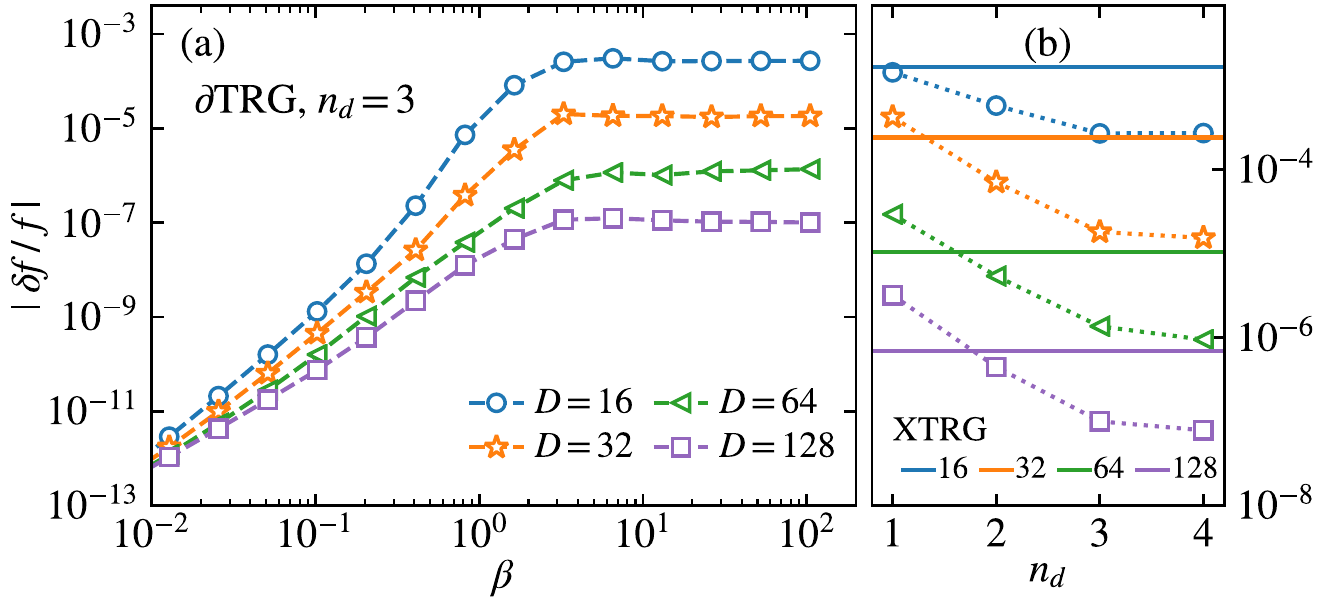}
\caption{(Color online) (a) Relative errors of transverse-field Ising model on a $4\times4$ open square lattice, at critical transverse field $h=h_c$. (b) \dTRG results with various $D$ are plotted versus optimization depths $n_d=1,2,3$ and 4. The comparison is at a fixed low temperature $\beta\simeq105$, and the standard XTRG results are shown with solid lines.
}
\label{Fig:4X4}
\end{figure}

\begin{figure}[!tbh]
\includegraphics[angle=0,width=1\linewidth]{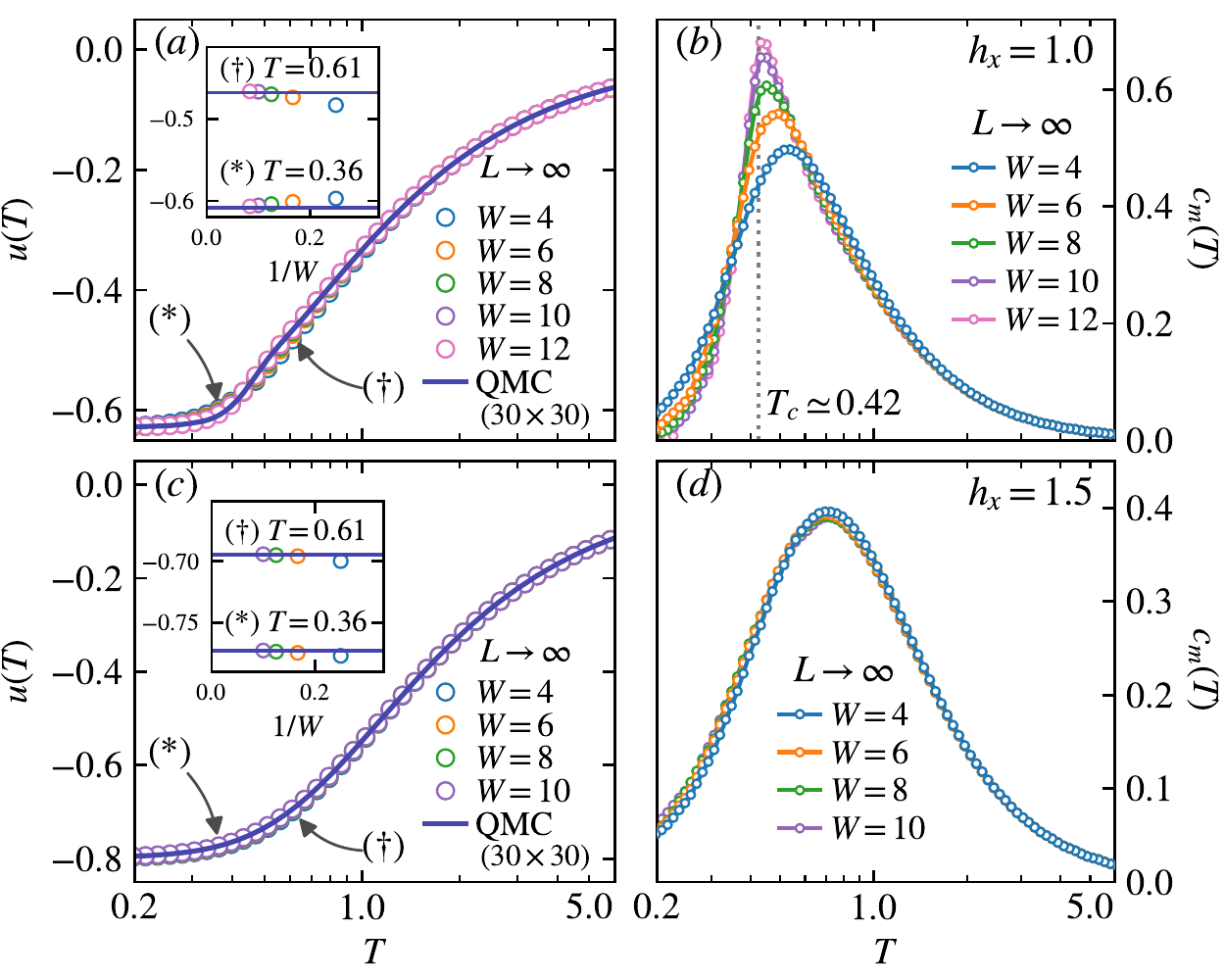}
\caption{(Color online)
{The plot show internal energy $u(T)$ and spin specific heat $c_m(T)$ of transverse field Ising model, with $h_x=1$ [(a, b)]  and $h_x=1.5$ [(c, d)].}
$\partial$TRG calculations are run on cylinders of various widths $W$ (with length $L$ extrapolated to infinity). 
{The energy data show excellent agreements with the quantum Monte Carlo (QMC) result 
\cite{Bauer.b+:2011:ALPS} on a $30\times30$ square lattice  in both fields. 
For the $h_x=1$ case, the peak position in $c_m(T)$ provides an accurate estimate, 
$\sim1$\% relative error, of critical temperature.}
}
\label{Fig:Inf}
\end{figure}

\textit{Thermodynamics of finite-size quantum lattice models.---} 
Now we apply $\partial$TRG to finite-size chains and cylindrical geometries of finite width $W$ (and length $L$), 
and try to  approach the thermodynamics limit by increasing the system size, {which has been proved to be very successful in ground-state simulations of quantum frustrated magnets \cite{White2007,He2017}}.
Note that the sweep optimization needs to be adapted when applied to the finite-size systems, i.e., 
we not only scan $\llb{w^{(i)}}$ between different scales but also among different lattice sites/bonds. 

For a finite-size system on the 1D or 2D lattice, the high-$T$ density matrix $\rho(\tau)$ can be initialized 
through a discretization-error-free series expansion technique \cite{Chen.b+:2017:SETTN, SM}. 
It has been shown to be preferable, over Trotter-Suzuki type initializations, in dealing 2D systems defined on, 
e.g., long cylinders \cite{Chen2018,Li2019}. The benchmark results on the finite-size 
XY chain can be found in Supplementary Materials \cite{SM}, 
where deep optimization into $n_d$ layers gains remarkable improvement in accuracy.

As a demonstration for 2D simulations, below we focus on the transverse-field Ising model on a square lattice, i.e.,  
\begin{equation}
H = \sum_{\langle i,j \rangle} J S_i^z S_j^z - {h_x} \sum_i S_i^x,
\label{Eq:TFI}
\end{equation} 
where $J=-1$ (ferromagnetic) is set as the energy scale. The model undergoes a magnetic order-disorder quantum phase transition at a critical field $h_c\simeq 1.52219(1)$ \cite{Bloete2002}. Through a snake-path mapping into quasi-1D lattice \cite{Li2019}, the interaction information of the Hamiltonian Eq.~(\ref{Eq:TFI}) on a width $W$ cylinder can be encoded in a compact MPO of bond dimension $D_H = W +2$ \cite{Chen2018,Li2019b}. Given the MPO representation of $H$, \dTRG works automatically and produces accurate results as benchmarked below.

Firstly, we run \dTRG simulations on a 4$\times$4 square lattice at the critical transverse field $h=h_c$, and compare the results to exact diagonalization data. In \Fig{Fig:4X4}(a), the relative errors $|\delta f/f|$ are plotted vs. $\beta$. One can observe a high accuracy with an optimization depth $n_d=3$, which continuously improves upon increasing the bond dimension $D$. Moreover, to reveal the effects of $n_d$, in \Fig{Fig:4X4}(b) we show $|\delta f/f|$ vs. $n_d$ at low temperature, as compared to XTRG data. Indeed the accuracy improves considerably, by orders of magnitude, as $n_d$ increases. For example, the $D=64,n_d=4$ \dTRG accuracy even goes parallel with the $D=128$ XTRG one. 

\textit{Large-scale simulations and finite-temperature phase transition.---} Next, we conduct \dTRG calculations of quantum Ising model on cylinders with various widths $W$ (up to 12) and lengths $L$. The transverse field is first fixed at $h_x=1.0 \approx 2/3h_c$, giving rise to a spin order in low temperature. The long-range order melts at a critical temperature $T_c\simeq 0.42$, through a second-order phase transition \cite{Czarnik.p+:2015:PEPS}. In \Fig{Fig:Inf}, we retain only moderate bond dimension up to $D=128$ in the calculations, and the optimization depth is up to $n_d=4$ layers. The internal energy $u(T)$ and magnetic specific heat $c_m(T)$ are computed from the first and second numerical derivatives of $f(T)$, respectively. Following the line developed in XTRG \cite{Chen2018,Li2019}, we exploit a $z$-shift technique as well as numerical interpolation to collect dense enough data points and ensure a negligible differential error \cite{SM}. Moreover, to eliminate the finite-length effects, we perform an extrapolation to $L=\infty$, via linear fitting or energy subtraction \cite{SM}. 

Collecting the extrapolated data at each width $W$, we compare the internal energy $u(T)$ with QMC data in \Fig{Fig:Inf}(a). A very good agreement of our cylindrical results with the large-scale QMC data is obtained. The latter is computed on a $30\times30$ square lattice with periodic boundary condition (i.e., torus) which mimics the thermodynamic limit. Furthermore, as shown in the inset, we zoom in at two selected temperatures and find there relative errors $|\delta u/u|\sim10^{-3}$ ($W=12$ result), with respect to QMC. In \Fig{Fig:Inf}(b), by further taking the derivatives of internal energy $u(T)$, we obtain the specific heat curves $c_m(T)$. It is observed that the peak in $c_m(T)$ gets sharper as $W$ increases, signaling the existence of a phase transition, and the peak locations for wide cylinders are in very good agreement with $T_c\simeq0.42$ in the thermodynamic limit.

In \Fig{Fig:Inf}(c,d), we provide $u(T)$ and $c_m(T)$ at transverse field $h_x=1.5\simeq 0.99h_c$, in close vicinity of the quantum phase transition point. Again, $u(T)$ results are in excellent agreement ($|\delta u/u|\sim10^{-4}$) with the QMC data as shown in \Fig{Fig:Inf}(c). It is observed in \Fig{Fig:Inf}(d) that the specific heat shows a round peak at around $T/J=0.7$, which has well converged vs. system sizes and does not correspond to any phase transition, which should occur at below $T/J=0.2$.

\textit{Conclusion and outlook.---} 
Inspired by the essential correspondence between the backpropagation and SRG of tensor networks, we propose the framework of $\partial$TRG. With \dTRG, we make much better use of tensor parameters by increasing the optimization depth, instead of merely enlarging parameter space dimension $D$. As a result, a moderate $D$ can lead to an unprecedented high accuracy in simulating thermodynamics of 2D quantum models.
Bearing the virtue of SRG, it can optimize both the wave function representation and the renormalization transformations, globally and automatically.
 Therefore, \dTRG constitutes a promising tool to investigate very challenging manybody problems, e.g., frustrated antiferromagnets, fermionic Hubbard models, which are currently of great research interest.

\textit{Acknowledgments}.--- We are indebted to Jian Cui, Jan von Delft, Yannick Meurice, and Andreas Weichselbaum for helpful discussions. This work was supported by the National Natural Science Foundation of China (Grant Nos. 11774420, 11834014, 11974036, and 11774398), the National R$\&$D Program of China (Grants Nos. 2016YFA0300503, 2017YFA0302900), German Research Foundation (DFG WE4819/3-1) under Germany's Excellence Strategy\,-\,EXC-2111\,-\,390814868 and by the Research Funds of Renmin University of China (Grants No. 20XNLG19).
Our code implementation in PyTorch is publicly available at this \href{https://github.com/TensorBFS/dTRG}{https URL}.

\AtEndEnvironment{thebibliography}{
\bibitem{footnote} Here for the sake of computational cost, we exploit the MERA update for 
$w^{(i)}$ at the first scale after the environment is obtained through the backward process, 
while the rest $w^{(i)}$ on higher layers are update via HOTRG technique.
\bibitem{pytorch} The official website of PyTorch is at \href{https://pytorch.org/}
{https://pytorch.org}.
\bibitem{autograd} See, e.g., \href{https://github.com/HIPS/autograd}{https://github.com/HIPS/autograd}.
\bibitem{SM} In Supplemental Material, we briefly recapitulate the basic idea of TRG in 
{\color{blue}Sec.~A}, SRG in {\color{blue}Sec.~B}, backpropagation in {\color{blue}Sec.~C}, 
and the Quasi-Newton optimization of isometries in {\color{blue}Sec.~D}. Detailed discussions 
on the correspondence between SRG and backpropagation are provided in {\color{blue}Sec.~E}, 
and the initialization of $\rho(\tau)$ in {\color{blue}Sec.~F}. Besides, the analytical solutions 
({\color{blue}Sec.~G}), computational hours in infinite XY chain ({\color{blue}Sec.~H}), 
finite XY chain results ({\color{blue}Sec.~I}), the $z$-shift technique ({\color{blue}Sec.~J}), 
and the energy extrapolation in 2D transverse-field Ising model ({\color{blue}Sec.~K}) are 
also presented.
}

\bibliography{Reference}

\newpage
\mbox{}

\begin{center}
\textbf{\large Supplemental Materials: Automatic Differentiation for Second Renormalization of Tensor Networks}
\end{center}

\maketitle

\setcounter{section}{0}
\setcounter{figure}{0}
\setcounter{equation}{0}
\renewcommand{\thesection}{\Alph{section}}
\renewcommand{\theequation}{S\arabic{equation}}
\renewcommand{\thefigure}{S\arabic{figure}}
\renewcommand{\theequation}{\Alph{section}\arabic{equation}}

\section{Tensor network representation of the Ising model and its renormalization}
\label{App:TNIsing}
In this section, we briefly discuss some basic notions on the real-space tensor renormalization group (TRG) methods, which was referred to as \textit{rewiring method} in literatures, e.g., Ref.~\cite{Kadanoff-RMP2014}. For the sake of simplicity, we take the Ising model on the square lattice as an example. Due to the locality of the interaction, the partition function has a compact tensor-network representation \cite{SRG-PRB2010}, i.e.,
\beq
Z = \sum_{\{\sigma\}}e^{-\beta H(\{\sigma\})} = \sum_{\{\sigma\}}\prod_{a} T_{\sigma_i\sigma_j\sigma_k\sigma_l}^{(a)}
\eeq
where $\{\sigma\}$ denotes the classical spin configurations in the Hamiltonian $H$. The tensor $T^{(a)}$ is defined on a local plaquette, labeled as $a$, containing the four original spins. The central idea of renormalization group lies in the concept of renormalization transformation \cite{Wilson-RMP1975, Kadanoff-PRL1975}, in which a set of new coupling constants $\{K\}$ is sought to represent the Hamiltonian as 
\beq
H = \sum_{a,i}K_is^{(a)}_i,  \label{RGtrans}
\eeq
where $s^{(a)}_i$ defined at a larger scale is a ``block spin", i.e., combination of original spins $\{\sigma\}$ in the plaquette $a$. 

Accordingly, in the rewiring method, TRG finds a set of tensors $\{T^{(a')}\}$ defined at plaquette $a'$ (thus at a larger length scale) to represent the partition function, i.e., 
$$Z = \sum_{\{\alpha\}}\prod_{a'} T_{\alpha_i\alpha_j\alpha_k\alpha_l}^{(a')},$$ 
where the tensors $T$ are obtained following a similar line of Kadanoff's RG transformation: We introduce new statistical variables $\{\alpha\}$ (as geometric indices of tensors) and then trace out the original variables $\{\sigma\}$. TRG methods provide accurate tool and versatile platform for studying the conformal criticality and universality near phase transition temperatures \cite{Kadanoff-RMP2014}.

In TRG, the transformations between $T_{\sigma_i\sigma_j\sigma_k\sigma_l}$ at small length scale and $S_{\alpha_i\alpha_j\alpha_k\alpha_l}$ at a larger one constitute the most important parameters of the program. Once the transformations are given, they renormalize the system and result in the partition function $Z$. In most cases, approximations have to be introduced in the course of TRG, while causing minimal loss in the partition function. We point out that, while most TRG programs employs renormalization transformations that are found only locally \cite{Levin.m+:2007:TRG, TEFR-PRB2009, TNR-PRL2015, LoopTRG-PRL2017, PosTRG-PRL2017}, the second renormalization group (SRG) \cite{Xie2009, SRG-PRB2010, Xie2012, HOSRG-PRB2016} manages to find proper transformations that optimizes the partition function globally. 

\section{Second renormalization group}
\label{App:SRG}
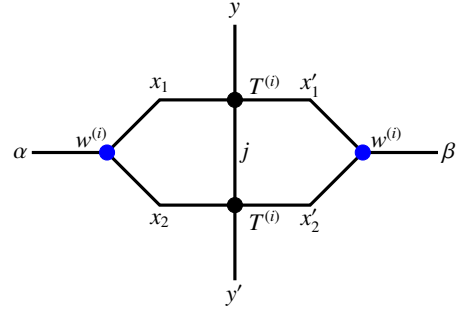
\begin{figure}
\centering
\begin{tikzpicture}
\draw[very thick] (0, 0) -- (1, 0) -- (1.7, 0.7) -- (3.7, 0.7) -- (4.4, 0) -- (5.4, 0);
\draw[very thick] (1, 0) -- (1.7, -0.7) -- (3.7, -0.7) -- (4.4, 0);
\draw[very thick] (2.7, 1.7) -- (2.7, -1.7);
\filldraw[blue] (1, 0) circle (0.1);
\filldraw[blue] (4.4, 0) circle (0.1);
\filldraw[black] (2.7, 0.7) circle (0.1);
\filldraw[black] (2.7, -0.7) circle (0.1);
\draw (-0.15, 0) node(X) {$\alpha$};
\draw (5.55, 0) node(X) {$\beta$};
\draw (1.7, 0.9) node(X) {$x_1$};
\draw (1.7, -0.9) node(X) {$x_2$};
\draw (3.7, 0.9) node(X) {$x'_1$};
\draw (3.7, -0.9) node(X) {$x'_2$};
\draw (2.7, 1.9) node(X) {$y$};
\draw (2.82, 0) node(X) {$j$};
\draw (2.7, -1.9) node(X) {$y'$};
\draw (3.1, 0.9) node(X) {$T^{(i)}$};
\draw (3.1, -0.9) node(X) {$T^{(i)}$};
\draw (0.8, 0.2) node(X) {$w^{(i)}$};
\draw (4.72, 0.2) node(X) {$w^{(i)}$};
\end{tikzpicture}
\caption{Illustration of forward process in Eq.~(\ref{HOTRGstep}). Note that in the backward recursive equation Eq.~(\ref{SRG}), the upper $T^{(i)}$ is denoted as $T^{(i;u)}$, and the lower one $T^{(i;d)}$. }
\label{Fig:HOTRG}
\end{figure}

In the following, we recapitulate SRG in the higher-order tensor renormalization group (HOTRG) algorithm, i.e., HOSRG \cite{Xie2012,HOSRG-PRB2016}. For a lattice system with translational invariance, $T^{(i)}$ is used to denote the local tensors renormalized at the $i$-th scale. $i=1$ denotes the initial scale at which the original Hamiltonian is defined, and $w^{(i)}$ is the $i$-th isometric renormalization transformation. 
As shown in Fig.~\ref{Fig:HOTRG}, given the renormalization transformations, we can perform a forward HOTRG iteration, 
\beq
T^{(i+1)}_{\alpha\beta yy'} = \sum_{jx_1x_2x'_1x'_2} T^{(i)}_{x_1x'_1yj}T^{(i)}_{x_2x'_2jy'} w^{(i)}_{x_1x_2\alpha} w^{(i)}_{x'_1x'_2\beta} \label{HOTRGstep}
\eeq
where the subscripts ${x_1, x_2, \alpha}$ of $w_{x_1x_2\alpha}^{(i)}$ can be understood as the new statistical variables introduced at different scales, with which the Hamiltonian can be written down, c.f., Eq.~(\ref{RGtrans}).
The transformations $w^{(i)}$ between  two scales $i$ and $i+1$ are determined by the local higher-order singular value decompositions according to tensors $T^{(i)}$. 

In the forward HOTRG iteration described above, the transformations $w^{(i)}$ is determined locally \cite{Xie2012}. To achieve a global optimization, the environment tensor $E$ should be considered. We punch a ``hole'' by removing the target tensor in the tensor network and contract all the indices except those of the target. A backward iteration should be involved to accomplish this task and perform the global optimization. 

For example, the environment of $T^{(i)}$ at the target scale $i$ can be obtained exploiting the following recursive relation
\beqn
E^{(i;u)}_{x_1x'_1yj} = \sum_{\alpha\beta y'x_2x'_2}E^{(i+1)}_{\alpha\beta yy'}T^{(i;d)}_{x_2x'_2jy'}w^{(i)}_{x_1x_2\alpha}w^{(i)}_{x'_1x'_2\beta} \label{SRG}
\eeqn
where the superscripts $\lambda = \lsb{u,d}$ denote the upper and lower $T^{(i)}$ in Fig.~\ref{Fig:HOTRG}, and environment $E^{(i)}$ is averaged over two equivalent environment, i.e., $E^{(i)} = \sum_{\lambda}E^{(i;\lambda)} / 2$, with $\lambda=u,d$. 
Given the environment tensor $E^{(i)}$, we can update the renormalization transformation $w^{(i)}$ that globally optimizes the partition function.

\section{Backpropagation in neural networks}
\label{App:Backprop}
Generally speaking, a deep feedforward neural network sets up a mapping between
a set of input signals $x^{(1)}$, such as images, and a set of output signals $y$, say, categories, through a multi-layer transformation $\mathcal{F}$, i.e., $y = \mathcal{F}(x^{(1)})$, where $\mathcal{F}$ is represented as a composition of many arranged linear ($\mathcal{L}$) transformations separated by nonlinear ($\mathcal{N}$) mappings. To be specific,
an $n$-layer neural network $\mathcal{F}$ can be expressed as
\beq
\mathcal{F} = \mathcal{N}^{(n)}\mathcal{L}^{(n)}...\mathcal{N}^{(2)}\mathcal{L}^{(2)}\mathcal{N}^{(1)}\mathcal{L}^{(1)}
\eeq
where the linear transformations $\mathcal{L}$'s contain most of the variational parameters $\omega$'s that need to be optimized. Roughly speaking, the major goal of training a neural network is to find the optimal $\omega$'s which minimize the target objective function $L$ characterizing the discrepancy between the actual and predicted labels. 

In essence, each layer of the neural network, i.e., $\mathcal{L}^{(i)}$ followed by $\mathcal{N}^{(i)}$, perform the transformations on the input data/feature $x^{(i)}$, and its output $x^{(i+1)}$ can be regarded as a new representation and a higher-level abstraction of original input \cite{SHLi2018,Koch-Janusz2018}. This is very similar to the renormalization transformation of TRG described in the main text, where new bases are introduced in different scales to represent the original partition function.

To train the neural network, the gradients of the objective function $L$ with respect to the parameters $\omega$'s are required for a global optimization. The backpropagation algorithm, arguably the most successful approach for training deep neural networks, exploits the chain rule of derivatives to efficiently compute these gradients \cite{Hinton-Nature1986, Parker-LLR1985, LeCun-IEEE1989,LeCun-Nature2015}. To be concrete, the backpropagation method relates the gradients in two neighboring layers as follows
\beq
\pd{L}{x^{(i)}} = \pd{L}{x^{(i+1)}}\cdot\pd{x^{(i+1)}}{x^{(i)}}.  \label{BP}
\eeq

In deep learning, automatic differentiation (AD) technique is employed to compute the derivative $\pd{L}{x^{(i)}}$ following the recursive equation \Eq{BP}, in an analytically rigorous way. Furthermore, the derivatives with respect to the parameters $\omega^{(i-1)}$ can be similarly obtained via one more AD step through $\pd{x^{(i)}}{\omega^{(i-1)}}$, with which one can update $\omega^{(i-1)}$ and regenerate $x^{(i)}$ in the $i$-th layer.

\section{Quasi-Newton optimization of isometries}
\label{App:LBFGS}
Given the gradient $E_{w}^{(i)} =\frac{1}{N_i} \overline{w^{(i)}}$ [see Eq.~({\color{blue}1}) in the main text], the quasi-Newton approach constitutes a class of efficient algorithm for parameter optimization. Here in \dTRG, to impose the isometric constraint, we choose the parametrization $w^{(i)} =  \exp{\lsb{B^{(i)}}}$, where $B^{(i)}$ is an anti-symmetric real square matrix of dimension $D^2$. 

With this parameterization, the standard gradient based L-BFGS method \cite{Wright-NO2006} can be employed to update $B^{(i)}$ so as to maximize $Z$. After an exponential operation, we can then compute $\exp{\lsb{B^{(i)}}}$ numerically and then truncate this unitary matrix (of dimension $D^2$, see main text) into a $D^2\times D$ isometry $w^{(i)}$.

Note the computational cost of this ``tailored" quasi-Newton approach  is of time complexity $O(D^6)$, higher than the MERA update $O(D^4)$ that we used in the main text.

\section{Details on the correspondence between SRG and backpropagation}
\label{App:Equiv}
Suppose the tensor elements in $w^{(i)}$s are independent variational parameters, we can bridge SRG and backpropagation approach, in the context of  higher-order tensor renormalizations, by comparing Eq.~(\ref{SRG}) and Eq.~(\ref{BP}). 

\textit{From SRG to backpropagation}: Firstly, we assert the following formula which equals ``punching a hole" in the tensor network to the derivative
\beqn
E^{(i;\lambda)}_{abcd} &=& \frac{1}{N_i}\cdot\pd{Z}{T^{(i;\lambda)}_{abcd}} = \frac{\sum_{j}E^{(i,j;\lambda)}_{abcd}}{N_i},  \label{Eeq}
\eeqn
where $2N_i = 2^{(n-i+1)}$ is the number of $T^{(i)}$ copies (labeled by index $j$) in the partition function tensor network, i.e., $Z = \mathrm{Tr}\lsb{T^{(i)}}^{\otimes 2N_i}$. Matter of fact, the equation $$Z = \sum_{abcd}T^{(i,j;\lambda)}_{abcd}E^{(i,j;\lambda)}_{abcd}$$ holds at all scales $i$ and location $j$. Next, we point out that the tensor contractions in Eq.~(\ref{SRG}) is nothing but multiplying Jacobian to the tensor $E^{(i+1)}$, i.e.,
\beqn
\pd{T^{(i+1)}_{\alpha\beta yy'}}{T^{(i;u)}_{x_1x'_1zj}} &=& \delta_{yz}\cdot\sum_{x_2x'_2}T^{(i;d)}_{x_2x'_2jy'}w^{(i)}_{x_1x_2\alpha}w^{(i)}_{x'_1x'_2\beta}, \label{Teq}
\eeqn
which is the derivative of $T^{(i+1)}$ with respect to $T^{(i;u/d)}$, self-evident in Fig.~\ref{Fig:HOTRG}. Combining Eq.~(\ref{Eeq}) and Eq.~(\ref{Teq}) together, we can rewritten Eq.~(\ref{SRG}) as
\beqn
\pd{Z}{T^{(i)}_{abcd}} &=& \sum_{xyzw}\pd{Z}{T^{(i+1)}_{xyzw}}\cdot\pd{T^{(i+1)}_{xyzw}}{T^{(i)}_{abcd}} \label{Unify}
\eeqn
where $\pd{T^{(i+1)}}{T^{(i)}} = \sum_{\lambda}\pd{T^{(i+1)}}{T^{(i;\lambda)}}$ is assumed.

Therefore, we reach the conclusion that the recursive relation of environment in the SRG backward iteration, as expressed in Eq.~(\ref{SRG}), is exactly the derivative chain rule Eq.~(\ref{BP}) in backpropagation method of deep learning. In Eq.~(\ref{Unify}), we have customized the backpropogation with the tensor network context to emphasize the in-depth link between the two.

Moreover, to show the correspondence in a more intuitive way, we introduce the following notations

\beqn
\begin{tikzpicture}[baseline = -0.5ex]
\draw[red, very thick] (0, 0) circle(0.6);
\draw[very thick] (-0.6, 0) -- (-0.3, 0)
 (0.6, 0) -- (0.3, 0)
 (0, 0.6) -- (0, 0.3)
 (0, -0.6) -- (0, -0.3);
\draw (-0.4, 0.15) node {$a$};
\draw (0.4, 0.2) node {$b$};
\draw (0.15, 0.4) node {$c$};
\draw (0.15, -0.4) node {$d$};
\end{tikzpicture}
&\doteq& E^{(i)}_{abcd} = \frac{1}{2N_i}\cdot\pd{Z}{T^{(i)}_{abcd}},\nonumber \\
\begin{tikzpicture}[baseline = -0.5ex]
\draw[green!70!white, very thick] (0, 0) circle(0.6);
\draw[very thick] (-0.6, 0) -- (-0.3, 0)
 (0.6, 0) -- (0.3, 0)
 (0, 0.6) -- (0, 0.3)
 (0, -0.6) -- (0, -0.3);
\draw (-0.4, 0.15) node {$x$};
\draw (0.4, 0.2) node {$y$};
\draw (0.15, 0.4) node {$z$};
\draw (0.15, -0.4) node {$w$};
\end{tikzpicture}
&\doteq& E^{(i+1)}_{xyzw} = \frac{1}{N_i}\cdot\pd{Z}{T^{(i+1)}_{xyzw}},   \label{Note1}
\eeqn
and represent the Jacobian $\pd{T^{(i+1)}}{T^{(i)}}$ in Eq.~(\ref{Teq}) as
\beq
\begin{tikzpicture}[baseline = -0.7ex]
\draw[very thick] (-0.95, 0) -- (-0.55, 0) -- (-0.4, -0.3) -- (0.4, -0.3) -- (0.55, 0) -- (0.95, 0)
(-0.55, 0) -- (-0.4, 0.3) -- (-0.1, 0.3)
(0.55, 0) -- (0.4, 0.3) -- (0.1, 0.3)
(0, 0.7) -- (0, 0.4)
(0, 0.2) -- (0, -0.7);
\draw (-0.75, 0.15) node {$x$};
\draw (0.75, 0.2) node {$y$};
\draw (0, 0.7) node[above] {$z=c$};
\draw (0, -0.7) node[below] {$w$};
\draw (-0.4, 0.45 ) node {$a$};
\draw (0.4, 0.5 ) node {$b$};
\draw (0.15, -0.05) node {$d$};
\filldraw[blue] (-0.55, 0) circle(0.1);
\filldraw[blue] (0.55, 0) circle(0.1);
\filldraw[black] (0, -0.3) circle(0.1);
\end{tikzpicture}
\squad + \squad
\begin{tikzpicture}[baseline = -0.7ex]
\draw[very thick] (-0.95, 0) -- (-0.55, 0) -- (-0.4, 0.3) -- (0.4, 0.3) -- (0.55, 0) -- (0.95, 0)
(-0.55, 0) -- (-0.4, -0.3) -- (-0.1, -0.3)
(0.55, 0) -- (0.4, -0.3) -- (0.1, -0.3)
(0, 0.7) -- (0, -0.2)
(0, -0.4) -- (0, -0.7);
\draw (-0.75, 0.15) node {$x$};
\draw (0.75, 0.2) node {$y$};
\draw (0, 0.7) node[above] {$z$};
\draw (0, -0.7) node[below] {$w=d$};
\draw (-0.4, -0.5 ) node {$a$};
\draw (0.4, -0.5 ) node {$b$};
\draw (0.15, 0.05) node {$c$};
\filldraw[blue] (-0.55, 0) circle(0.1);
\filldraw[blue] (0.55, 0) circle(0.1);
\filldraw[black] (0, 0.3) circle(0.1);
\end{tikzpicture}
\squad \doteq \squad \pd{T^{(i+1)}_{xyzw}}{T^{(i)}_{abcd}}.    \label{Note2}
\eeq
Then we can picturize Eq.~(\ref{Unify}) as
\beqn
\begin{tikzpicture}[baseline = -0.7ex]
\draw[red, very thick] (0, 0) circle(0.4);
\draw[very thick] (-0.4, 0) -- (-0.2, 0)
 (0.4, 0) -- (0.2, 0)
 (0, 0.4) -- (0, 0.2)
 (0, -0.4) -- (0, -0.2);
\end{tikzpicture}
&=& \frac{1}{2}\times\lmb{
\begin{tikzpicture}[baseline = -0.7ex]
\draw[very thick] (-0.95, 0) -- (-0.55, 0) -- (-0.4, -0.3) -- (0.4, -0.3) -- (0.55, 0) -- (0.95, 0)
(-0.55, 0) -- (-0.4, 0.3) -- (-0.1, 0.3)
(0.55, 0) -- (0.4, 0.3) -- (0.1, 0.3)
(0, 0.7) -- (0, 0.4)
(0, 0.2) -- (0, -0.7);
\filldraw[blue] (-0.55, 0) circle(0.1);
\filldraw[blue] (0.55, 0) circle(0.1);
\filldraw[black] (0, -0.3) circle(0.1);
\draw[green!70!white, very thick] (0, 0) ellipse(0.95 and 0.7);
\end{tikzpicture}
\squad + \squad
\begin{tikzpicture}[baseline = -0.7ex]
\draw[very thick] (-0.95, 0) -- (-0.55, 0) -- (-0.4, 0.3) -- (0.4, 0.3) -- (0.55, 0) -- (0.95, 0)
(-0.55, 0) -- (-0.4, -0.3) -- (-0.1, -0.3)
(0.55, 0) -- (0.4, -0.3) -- (0.1, -0.3)
(0, 0.7) -- (0, -0.2)
(0, -0.4) -- (0, -0.7);
\filldraw[blue] (-0.55, 0) circle(0.1);
\filldraw[blue] (0.55, 0) circle(0.1);
\filldraw[black] (0, 0.3) circle(0.1);
\draw[green!70!white, very thick] (0, 0) ellipse(0.95 and 0.7);
\end{tikzpicture} } \nonumber \\
&=& \frac{1}{2}\times
\begin{tikzpicture}[baseline = -0.7ex]
\draw[green!70!white, very thick] (0, 0) circle(0.4);
\draw[very thick] (-0.4, 0) -- (-0.2, 0)
 (0.4, 0) -- (0.2, 0)
 (0, 0.4) -- (0, 0.2)
 (0, -0.4) -- (0, -0.2);
\end{tikzpicture}
\times \lmb{
\begin{tikzpicture}[baseline = -0.7ex]
\draw[very thick] (-0.95, 0) -- (-0.55, 0) -- (-0.4, -0.3) -- (0.4, -0.3) -- (0.55, 0) -- (0.95, 0)
(-0.55, 0) -- (-0.4, 0.3) -- (-0.1, 0.3)
(0.55, 0) -- (0.4, 0.3) -- (0.1, 0.3)
(0, 0.7) -- (0, 0.4)
(0, 0.2) -- (0, -0.7);
\filldraw[blue] (-0.55, 0) circle(0.1);
\filldraw[blue] (0.55, 0) circle(0.1);
\filldraw[black] (0, -0.3) circle(0.1);
\end{tikzpicture}
\squad + \squad
\begin{tikzpicture}[baseline = -0.7ex]
\draw[very thick] (-0.95, 0) -- (-0.55, 0) -- (-0.4, 0.3) -- (0.4, 0.3) -- (0.55, 0) -- (0.95, 0)
(-0.55, 0) -- (-0.4, -0.3) -- (-0.1, -0.3)
(0.55, 0) -- (0.4, -0.3) -- (0.1, -0.3)
(0, 0.7) -- (0, -0.2)
(0, -0.4) -- (0, -0.7);
\filldraw[blue] (-0.55, 0) circle(0.1);
\filldraw[blue] (0.55, 0) circle(0.1);
\filldraw[black] (0, 0.3) circle(0.1);
\end{tikzpicture} } \nonumber \\   \label{Note3}
\eeqn
where the open indices represent identities, same as those in Eq.~(\ref{Note1}) and Eq.~(\ref{Note2}). 

\textit{From backpropagation to SRG}: Now we start from \Eq{Unify} and try to recover the SRG operation. Given that the tensor network derivative equals environment tensor $E^{(i)}$ (up to a factor), the key step in backpropagation approach is also the computing of Jacobian ${\partial T^{(i+1)}}/{\partial T^{i}}$. In AD technique, this is implicitly expressed as a sequence of tensor contractions exactly reverse the forward process, as also shown in \Eq{Note2} (while from right hand side to left). That is to say, the Jacobian is computed in AD by contracting $T^{(i;d)}_{x_2x'_2jy'}, w^{(i)}_{x_1x_2\alpha}, w^{(i)}_{x'_1x'_2\beta}$ with the derivative ${\partial Z}/{\partial T^{(i+1)}} = N_{i+1} E^{(i+1)}$ from the $(i+1)$-th layer. Therefore, we again arrive at Eq.~(\ref{Note3}) and confirm that the recursive relation used in SRG is equivalent to the chain-rule AD procedure in backpropagation.

\section{Initialization of $\rho(\tau)$}
\label{App:Init}
In the simulation of quantum lattice models, we start from high temperature density operator $\rho(\tau)$, with a given manybody Hamiltonian $H$. There are two ways preparing $\rho(\tau)$ and obtaining its matrix product operator (MPO) representation. For infinite 1D quantum chains, we can employ a Trotter-Suzuki decomposition of $\rho(\tau)=e^{-\tau H}$, while for an finite-size system the series expansion technique offers us a discretization-error-free approach to prepare the MPO $\rho(\tau)$. Given the initial $\rho(\tau)$, we can perform successively the exponential cooling procedure down to the require low temperature. 

To perform the Trotter-Suzuki decomposition, we rewrite the Hamiltonian 
\begin{equation}
H_{XY} = \sum_{i} h_{i,i+1} = \sum_i S_i^x S_{i+1}^x + S_i^y S_{i+1}^y, 
\end{equation}
as
$$H_{XY} = H_o +H_e,$$ where $H_{o(e)}  = \sum_{i \in odd(even)} h_{i,i+1}$ contains odd(even) terms.
Therefore, up to $O(\tau^3)$ Trotter error, $e^{-\tau H_{XY}} = e^{-\frac{\tau}{2} H_o} e^{-\tau H_e} e^{-\frac{\tau}{2} H_o}$. With sufficiently small $\tau$, 
e.g., $\tau\simeq 5\times10^{-5}$ in Fig.~{\color{blue}3} of the main text, the Trotter error has been very well-controlled in practice.

On the other hand, for finite-size systems, including 2D systems mapped into quasi-1D chains with ``long-range" interactions,
we employ the series expansion of the density matrix
\begin{equation}
e^{-\tau H} = \sum_{n=1}^{N_c} \frac{ (-\tau)^n }{ n! } H^n,
\label{Eq:SETTN}
\end{equation} 
to realize an initialization of $\rho(\tau)$. By retaining sufficient large $N_c$, Eq.~(\ref{Eq:SETTN}) is free of any essential expansion error. 
Therefore, given an MPO representation of $H$, the density matrix $\rho(\tau)$ 
can be computed via the series-expansion machinery \cite{Chen.b+:2017:SETTN}, for both 1D and 2D finite-size systems.

Given the MPO representation of initial $\rho_0(\tau)$ at high temperature, 
the system can be cooled down linearly (LTRG) \cite{Li.w+:2011:LTRG} 
or exponentially  (XTRG) \cite{Czarnik.p+:2015:PEPS,Chen2018} 
along the $\beta$ axis. 
Due to the much fewer truncation steps,
it has been shown that XTRG constitutes a more accurate way of thermodynamic simulations \cite{Chen2018,Chen2019,Li2019}
and is thus adopted in the current work of \dTRG.

\begin{figure}[!tbp]
\includegraphics[angle=0,width=0.85\linewidth]{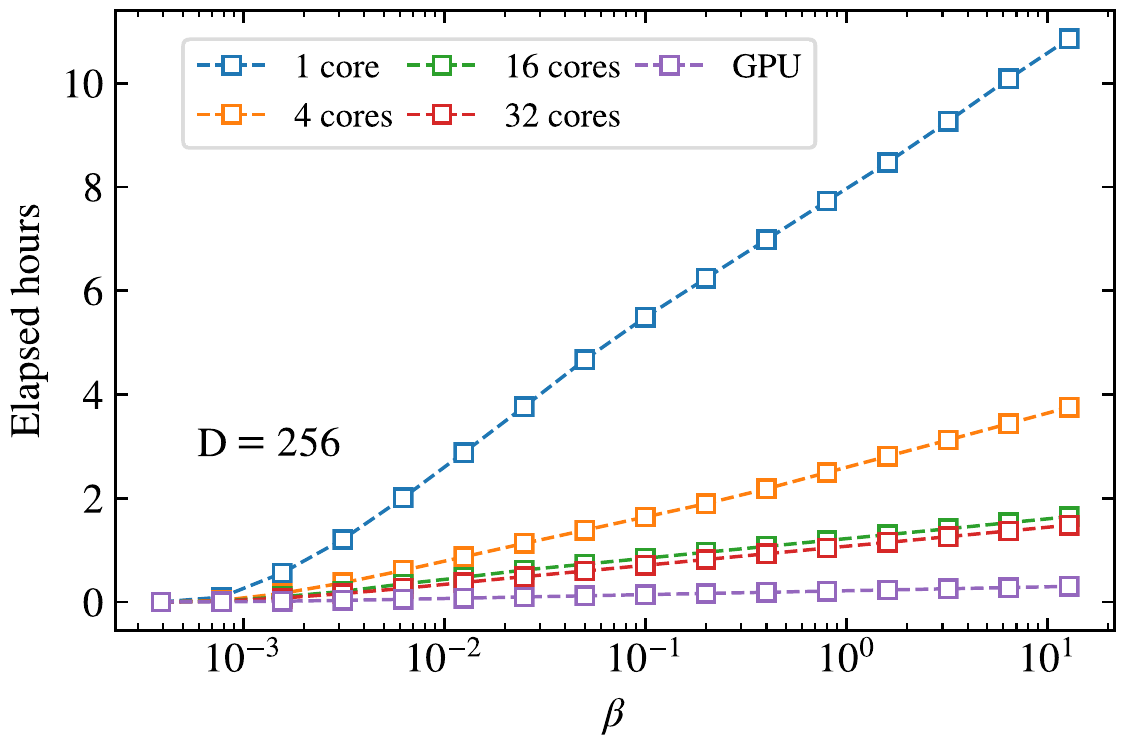}
\caption{(Color online) Elapsed hours $t_h$ scaling versus $\beta$, where a logarithmic scaling, i.e., $t_h \sim \ln{\beta}$ can be seen in both GPU and CPU (1, 4, and 16 cores) runs.  
}
\label{Fig:LogScaling}
\end{figure}

{
\section{Exact solution of quantum XY chain at finite temperature}
\label{App:Analytical}
We hereby provide the exact expression of partition function for 1-D quantum XY chain, 
\begin{equation}
H = J\sum_{i=1}^N \left(S^x_i S^x_{i+1} + S^y_i S^y_{i+1} \right) = \frac{J}{2}  \sum_{i=1}^N \left(S^+_i S^-_{i+1} + S^-_i S^+_{i+1}\right)
\end{equation}
with the periodic boundary condition $S^\pm_{N+1} = S^\pm_1$. 
Exploiting the Jordan-Wigner transformation
\begin{equation}
    \begin{cases}
S^+_j &= e^{-i\pi\sum_{k<j} c^\dagger_k c_k} c_j^\dagger, \\
S^-_j &= e^{i\pi\sum_{k<j} c^\dagger_k c_k} c_j^\dagger, \\
S^z_j &= c_j^\dagger c_j - \frac{1}{2},
    \end{cases}       
\end{equation}
the Hamiltonian can be expressed as a
spinless fermionic tight-binding chain,
\begin{equation}
H = \frac{J}{2}\left(\sum_{i=1}^{N-1} c^\dagger_i c_{i+1} - Q c^\dagger_N c_1\right) + h.c.
\end{equation}
with the parity $Q \equiv e^{-i\pi\sum_{j=1}^N c^\dagger_j c_j}$ being a conserved quantity. 
The Hilbert space then splits into two independent sectors: $Q=1$ (even particle number) sector; $Q=-1$, (odd particle number )sector, and the Hamiltonian can be expressed as, 
\begin{equation}
H = \frac{J}{2}\sum_{i=1}^{N} c^\dagger_i c_{i+1} + h.c.,
\end{equation}
with periodic(anti-periodic) boundary condition $c_{N+1} = \pm c_1$ in even(odd) sector. Through Fourier transformation $c_j = 1/\sqrt{N} \sum_q e^{-iqj} c_q$, the Hamiltonian gets diagonalized into, 
\begin{equation}
H = J\sum_q \cos{(q)} ~c_q^\dagger c_q = \sum_q \epsilon_q ~c_q^\dagger c_q,
\end{equation}
where $k$'s are summed over  modes $q=q_+\equiv 2\pi n/N$ in even sector while over $q=q_-\equiv 2\pi (n+1/2)/N$  in odd sector, with $n=0,1,\cdots,(N-1)$, to cope with corresponding boundary conditions. 

With even particle number constraint in $Q=1$ sector, the many-body states with odd numbers of modes should be excluded, when calculating the partition function \cite{Chen-CFT2017,Tu-CFT2017}. Thus, the partition function $Z_+$ in even sector is, 
\begin{equation}
Z_+(\beta) = \frac{1}{2}\prod_{q_+} (1+e^{-\beta\epsilon_{q_+}}) + \frac{1}{2}\prod_{q_+} (1-e^{-\beta\epsilon_{q_+}}).
\end{equation}
Similarly, the many-body states with even numbers of modes occupied should be excluded in $Q=-1$ sector, and the partition function reads as,
\begin{equation}
Z_-(\beta) = \frac{1}{2}\prod_{q_-} (1+e^{-\beta\epsilon_{q_-}}) - \frac{1}{2}\prod_{q_-} (1-e^{-\beta\epsilon_{q_-}}).
\end{equation}
Finally, one arrives at the partition function in the entire Hilbert space as, 
\begin{equation}
Z = Z_+ + Z_-,
\end{equation}
from which one can calculate the free energy and other thermodynamic quantities.
}

\section{XY chain: computational hours $t_h$ versus $\beta$}
\label{App:thscaling}
Here we provide the elapsed real time $t_h$ vs. $\beta$ in simulating the infinite XY chain. 
In Fig.~\ref{Fig:LogScaling}, we show the $D=256$ runs with various numbers of CPU cores (Xeon Gold 6230) 
as well as on the GPU (Tesla V100). We can see clearly a logarithmic scaling between
$t_h$ and $\beta$, for $\beta \gtrsim 0.1$. This logarithmic instead of linear scaling of wall time $t_h$ 
vs. $\beta$ clearly indicates the exponential speed up in \dTRG. In addition, from Fig.~\ref{Fig:LogScaling} we can see that in all cases 
(either GPU or CPU computations with various cores) $\beta=12.8$, 
the temperature point we have selected in Fig.~{\color{blue}3}(b) of the main text, 
is located well in the logarithmic regime.
That is to say, it constitutes a well suitable sampling temperature point 
for checking the $t_h$ vs. $D$ scaling in Fig.~{\color{blue}3}(b).

\section{$\partial$TRG calculations of finite-size XY chain}
\label{App:FXYC}

Due to the absence of translational invariance in finite-size systems, 
when applying $\partial$TRG to such systems, the isometries $w$ are bond dependent. 
Therefore, extra care is required in the optimization of $w$ tensors, as will be elaborated below.

Following that introduced in \Sec{App:Init}, we prepare the matrix product operator (MPO) 
representation of $\rho(\tau)$ via the series expansion. 
After the initialization, similar to the infinite cases, one perform iteratively renormalization of tensors 
to cool down the system from high to low temperatures, and can also sweep 
into inner $n_d$ layers. Nevertheless, there exist in finite-size \dTRG algorithms 
bond-dependent isometries to be optimized, thus sweeps amongst different bonds are required, 
along with those between different temperature scales.

\begin{figure}[!tbp]
\includegraphics[angle=0,width=0.86\linewidth]{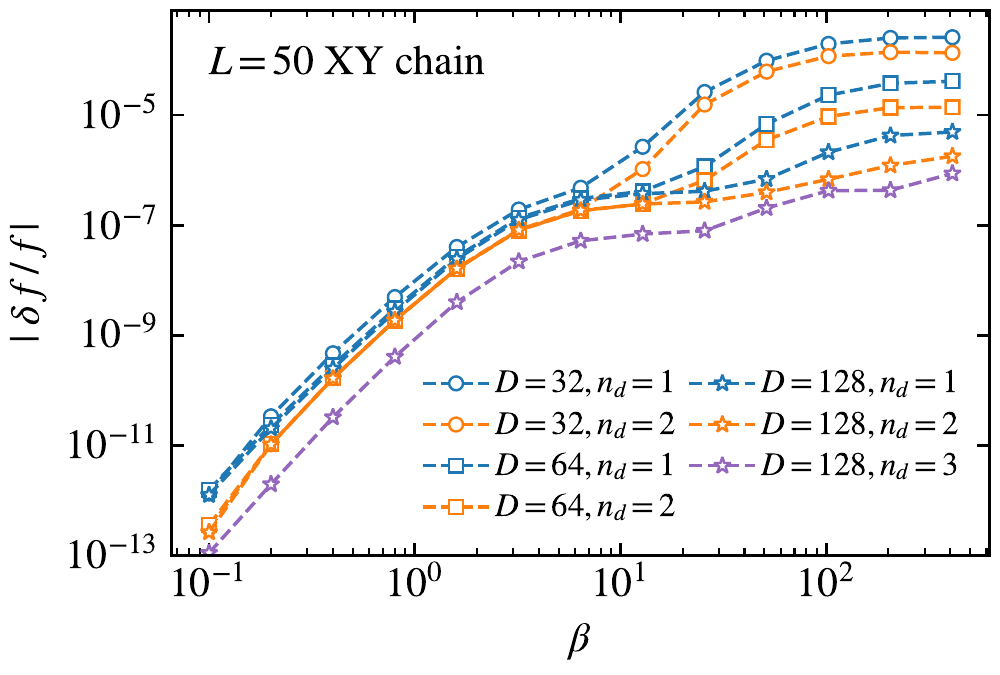}
\caption{(Color online) Relative errors of free energy $|\delta f/f|$ in $L=50$ quantum XY chain computed by 
$\partial$TRG. There is continuous improvement in the accuracies with the increase of bond dimensions $D=32,64$, and $128$,
as well as the sweep depths $n_d=1,2$, and $3$.}
\label{Fig:FiniteXYC}
\end{figure}

The finite-size XY spin chain can be solved exactly by a Jordan-Wigner transformation 
that maps the system into a non-interacting spinless fermion chain, 
from which the partition function can be readily obtained \cite{Chen2018}. 
In \Fig{Fig:FiniteXYC}, we perform the calculation of an $L=50$ XY chain and show the relative errors of free energy 
for various dimensions $D$ (up to $D=128$) and depths $n_d$ (up to 3).
Similar to the observations for infinite-size chain shown 
in Fig.~{\color{blue}3} of the main text, we can see in  \Fig{Fig:FiniteXYC} the accuracy 
improves significantly as the sweep depth $n_d$ increases.
Moreover, as shown in \Fig{Fig:FiniteXYC}, the improvement gets more and more 
pronounced as $D$ increases from $D=32$ to 128.
In particular, the improvement of accuracies gains over a wide range of temperatures, i.e., from high down to low temperatures. 
This again reveals unambiguously the advantage of deep optimization in \dTRG. 

\begin{figure}[!tbp]
\includegraphics[angle=0,width=0.99\linewidth]{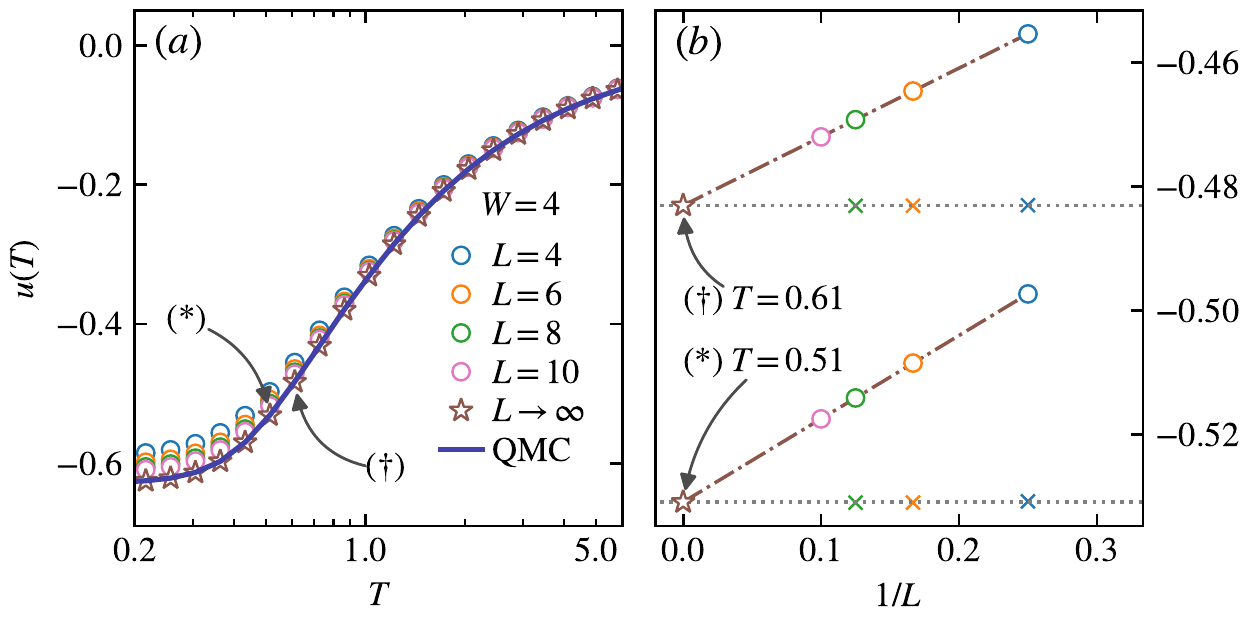}
\caption{(Color online) 
(a) Internal energy $u(T)$ of  TFI in field $h_x=1.0$ with fixed cylinder width $W=4$ and various $L$ (up to 10), which is used to extrapolate to $L=\infty$. It is benchmarked by QMC data with $W=4, L=\infty$ with similar extrapolations performed. 
(b) demonstrates the extrapolations through both the linear fitting and the subtraction technique (depicted as the cross marks), where excellent agreement is seen between the two schemes. The dotted horizontal line goes strictly through the extrapolated values (the star symbols), which is in perfect agreement with the subtraction results $u_{sub}$ (cross marks).
}
\label{Fig:extrap}
\end{figure}

\section{The $z$-shift technique}
\label{App:zshift}

In this appendix, we will briefly recapitulate the $z$-shift technique for the computation of thermodynamic quantities in \dTRG. Below we take the internal energy $u$ as an example, which is obtained by taking numerical derivative of free energy $f$, i.e., 
\begin{equation} \label{Eq:En}
u \equiv \pd{(\beta f)}{\beta} = \pd{(\beta f)}{\ln\beta}\cdot \frac{1}{\beta}
\end{equation}
as adopted in previous XTRG simulations \cite{Chen2018,Chen2019,Li2019}.
In the case of the temperature grid denoted as  
\begin{equation*}
\{\beta_i\} \equiv \{2^i\cdot\tau_0 \} = \{\tau_0, ~2\tau_0, ~4\tau_0, ~..., ~2^{n}\tau_0\}
\end{equation*}
being sparse, one can resort to the $z$-shift technique by shifting the initial temperature $\tau$ by a 
$z$-factor 
\begin{equation}
\tau = z \cdot \tau_0,\quad\mathrm{with}~z\in[1,2),
\end{equation}
and thus obtain a new grid 
\begin{equation*}
\{z\cdot\beta_i\} \equiv \{2^i\cdot\tau \} = \{\tau, ~2\tau, ~4\tau, ~..., ~2^{n}\tau\}.
\end{equation*}

Note that following the new grid $\{z\cdot\beta_i\}$ the simulations can be performed in parallel to the original $\{\beta_i\}$ run, thus constituting a highly efficient approach in XTRG \cite{Chen2018} as well as \dTRG. 

To be specific, as shown in Fig.~{\color{blue}5} of main text, we conduct the \dTRG simulations by following 4 sets of temperature grids $\{z\cdot\beta_i\}$ with $z$-factor chosen to be $z = 2^{0}, 2^{\sfrac{1}{4}}, 2^{\sfrac{1}{2}}, 2^{\sfrac{3}{4}}$. Before taking the numerical derivative \Eq{Eq:En}, in practice we further employ an interpolation of free energy data to reach an even denser temperature grid, i.e., totally 16 sets with $z = 2^0, 2^{\sfrac{1}{16}}, ..., 2^{\sfrac{15}{16}}$, which turns out to essentially eliminate the differential errors.

\section{Energy extrapolations of 2D transverse-field Ising model}
\label{App:extrap}

In this section, we demonstrate the extrapolations of internal energy $u(T)$ in transverse-field Ising (TFI) model on the cylindrical square lattice, via both linear fitting and subtraction methods, as mentioned in the main text.
In \Fig{Fig:extrap}(a), we show the internal energy in a $W=4$ TFI model with various length $L=4, 6, 8, 10$. To some extent, they already show nice convergence with each other as well as to the large-scale QMC data. 
Nonetheless, one can still extrapolate further to $L=\infty$ limit to eliminate the small finite-length effects, by either linear fitting $u_L = u_{\infty} + b/L$ or energy subtractions, i.e., $e_{sub} = (e_{L+2}-e_{L})/2W$ representing the `bulk' energy. 
As shown in \Fig{Fig:extrap}(b), both schemes generate mutually consistent energies in the $L=\infty$ limit, as indicated by the horizontal grey dotted lines which goes through exactly the extrapolated value [the asterisk symbol in \Fig{Fig:extrap}(b)]. Remind that the subtracted energy values get converged much faster to the infinite length limit than linear extrapolation and thus constitutes a more efficient technique in practice for extracting bulk energy expectation values. 


\end{document}